\documentclass[12pt]{article}

\usepackage{amsmath}
\usepackage{graphicx}
\usepackage{amssymb}
\usepackage{amsfonts}
\usepackage{latexsym}

\DeclareMathOperator{\cx}{\square}
\def\beq{\begin{eqnarray}}
\def\eeq{\end{eqnarray}}

\def\al{\alpha}
\def\be{\beta}

\def\ga{\gamma}
\def\de{\delta}
\def\vp{\varepsilon}

\def\ka{\kappa}
\def\la{\lambda}
\def\na{\nabla}

\def\si{\sigma}

\def\ph{\varphi}

\def\Ga{\Gamma}
\def\De{\Delta}

\begin{document}

%%%%%%%%%%%%%%%%%%%%%%%%%%%%%%
\begin{center}

{\large\bf Integration of trace anomaly in $6D$}
\vskip 4mm

Fabricio M. Ferreira~$^{a,b}$\footnote{
E-mail address: fabricio.ferreira@ifsudestemg.edu.br}
\
and
\
Ilya L. Shapiro~$^{a,c}$\footnote{
E-mail address: shapiro@fisica.ufjf.br}

%%%%%%%%%%%%%%%%%%%%%%%%%%%%%
\vskip 4mm

{\sl
(a) \ Departamento de F\'{\i}sica, ICE,
Universidade Federal de Juiz de Fora
\\
Campus Universit\'{a}rio - Juiz de Fora, 36036-330, MG, Brazil}
\vskip 2mm

{\sl
(b) \ Instituto Federal de Educa\c c\~ao, Ci\^encia e Tecnologia
do Sudeste de Minas Gerais \\
IF Sudeste MG - Juiz de Fora, 36080-001, MG, Brazil}
\vskip 2mm

{\sl
(c) \ Tomsk State Pedagogical University and Tomsk State
University, Tomsk, Russia}
\vskip 2mm
\end{center}
%%%%%%%%%%%%%%%%%%%%%%%%%%%%%%
\vskip 6mm

\begin{quotation}
\noindent
\textbf{Abstract.} \
The trace anomaly in six-dimensional space is given by the
local terms which have six derivatives of the metric. We find
the effective action which is responsible for the anomaly. The
result is presented in non-local covariant form and also in the
local covariant form which employs two auxiliary scalar fields.
\vskip 3mm

%%%%%%%%%%%%%%%%%%%%%%%%%%%%
\vskip 2mm

%%%%%%%%%%%%%%%%%%%%%%%%%%%%
\vskip 2mm

\end{quotation}
%%%%%%%%%%%%%%%%%%%%%%%%%%%%

%%%%%%%%%%%%%%%%%%%%%%%%%%%%
%%%%%%%%%%%%%%%%%%%%%%%%%%%%
%%%%%%%%%%%%%%%%%%%%%%%%%%%%
\section{Introduction}

The interest to the higher-dimensional conformal theories is on rise
since the advent of string/M-theory and the discovery of holography
and AdS/CFT correspondence. It would be certainly useful to have
an explicit form of the vacuum effective action for the conformal
fields in dimensions $D$ higher than four. The simplest and
practically working procedure to derive such an effective action
is by integrating conformal anomaly. The two main examples of
such integration are Polyakov action in $D=2$ \cite{Polyakov81}
and Riegert-Fradkin-Tseytlin action in $D=4$ \cite{rie}. Both
proved to be fruitful instruments for various applications (see,
e.g., \cite{Duff-94} for a review). The same integration in  $D=6$
attracts a great deal of attention, but until now there were only
particular (albeit very interesting) results \cite{Osborn} (see
further references therein) which do not enable one to obtain
the anomaly-induced effective action in a closed form.

In this Letter we report on a complete solution of the problem. The
work is organized as follows. In Sect. ~\ref{gen} we briefly describe
the scheme of integration which can be applied in any dimension $D$.
As one can see there, the three necessary elements are conformal
operator (analog of Paneitz operator in $D=4$), modified topological
invariant and its conformal transformation and, finally, the integration
of surface terms. The part which requires the most significant efforts
is the search of modified topological invariant with the simplest
conformal property, and we have this problem solved for $D=6$.
The relevant building blocks of such an effective action in  $D=6$
are presented in Sect.~\ref{main}. Finally, in Sect.~\ref{Conc} we
draw our conclusions are describe some of the possible applications.

%%%%%%%%%%%%%%%%%%%%%%%%%%%
\section{General scheme of integrating anomaly}
\label{gen}

Let us briefly summarize the general scheme of integrating anomaly,
as it is described in the review paper \cite{PoImpo} for $D=4$. The
changes which are requested in higher even dimensions are not
relevant, regardless of the growth of technical difficulties.

The vacuum part of the trace anomaly in dimension
$D \geqslant 4$ can be always written as
\cite{DDI,Duff77,DeserSchwimmer}
\beq
T
\,=\,
\langle T_\mu^\mu \rangle
\,=\, c_r\,W^{r}_D \,+\,a\,E_D \,+\,\Xi_D\,,
\label{T}
\eeq
with the sum over $r$. Here $W^{r}_D$  are conformal invariant terms
(typically constructed from Weyl tensor). In $D=2$ there is no conformal
term, and in $D=4$ there is only one, the square of the Weyl tensor. In
$D=6$ there are three such terms, the explicit form can  be found in
\cite{LPP}. Furthermore,  $\Xi_D$ is a linear combination of the surface
terms, $\Xi_D=\sum \ga_k\chi_k$ in the corresponding dimension. The
explicit form of the relevant $\chi_k$ terms in $D=6$ will be given
below in Eq.~(\ref{chi}). Furthermore, $E_D$ is the integrand of the
topological term,
\beq
E_D
&=&
\vp^{\rho_1\cdots \rho_D}\,\vp^{\si_1\dots \si_D}
\,R_{\rho_1\si_1\rho_2\si_2}\cdots
\,R_{\rho_{D-1}\si_{D-1}\rho_D\si_D}\,.
\label{GB-D}
\eeq
The classification (\ref{T}) is a simple consequence of that the
anomaly comes from the one-loop divergences and the last satisfy
conformal Noether identity. It is easy to see that the terms which
satisfy this identity should belong to the mentioned three categories.

The numerical coefficients $\,a$, $c$ and $\ga_k$ depend on the
number of massless conformal fields of different spins. These
quantities have no real concern to us, because we will describe
a general solution valid for any values of $\,a$, $c$ and $\,\ga_k$.

Our purpose is to find the anomaly-induced effective action
$\Ga_{ind}$, such that
\beq
-\,\frac{2}{\sqrt{-g}}\,g_{\mu\nu}
\,\frac{\de \Ga_{ind}}{\de g_{\mu\nu}}
\,=\,T\,.
\label{indEA}
\eeq

As it was already mentioned, the integration of anomaly
requires a modified topological invariant
\beq
\tilde{E}_D &=&
E_D + \sum\limits_k \al_k \chi_k\,,
\label{ED}
\eeq
where the values of $\al_k$ are chosen to provide the special
conformal property of the new topological term. Namely, we
require that under the local conformal transformation
\beq
g_{\mu\nu} &=&
{\bar g}_{\mu\nu}\,e^{2\si(x)}
\label{conf}
\eeq
there should be
\beq
\sqrt{-g}\tilde{E}_{D}
&=&
\sqrt{-{\bar g}}\big(\bar{\tilde{E}}_{D}
+ \ka {\bar \De}_{D}\si\big)\,,
\label{conf top}
\eeq
where $\ka$ is a constant and $\De_D=\cx^{D/2}+\dots$
is the conformal operator acting on a conformally inert scalar.
For example, in $D=4$, the  formulas have the well-known form, with
$\De_4$ being the Paneitz operator \cite{Paneitz,rie}, $\ka=4$,
and the surface term in (\ref{ED}) is $\al_k \chi_k=-(2/3)\Box R$.
Some comment is in order. Of course, in $D=4$ the $\Box R$ is
the unique possible surface term, so this part is simple. However,
the coefficient $-2/3$ is a little bit mysterious, because it can be
established only by a direct calculation. The details can be found
in \cite{Stud}, where one can observe that the conformal
transformation of each $E_4$ and $\Box R$ is quite complicated.
Nevertheless, the particular combination with the mystic $-2/3$
cancels all terms of second, third and fourth orders in $\si$ and
the remaining linear term involves the conformal operator. Indeed,
we expect this symmetry in the general even $D$ case, that means
\beq
\sqrt{-g}\De_D\ph  =  \sqrt{-\bar{g}}\bar\De_D \bar\ph
\label{confOP}
\eeq
with $\ph = \bar\ph$ and all other quantities with bar are
constructed with the fiducial metric $\,{\bar g}_{\mu\nu}$.

In order to integrate the anomaly one needs the last element. Namely,
there should be a set of {\it local} metric-dependent Lagrangians
${\cal L}_i$, providing that with some coefficients $c_{ik}$ there
is an identity
\beq
-\,\frac{2}{\sqrt{-g}}\,g_{\mu\nu}
\,\frac{\de}{\de g_{\mu\nu}}\,
\sum\limits_i c_{ik} \int_x{\cal L}_i \,=\,\chi_k\,,
\label{confLoc}
\eeq
where $\int_x \, \equiv \,\int d^Dx \sqrt{-g}$,
for each of the surface term components in (\ref{T}). If the set
${\cal L}_i$ is found, the problem of solving (\ref{indEA}) is
reduced to integrating the first two terms in (\ref{T}). And it is
easy to see that this problem is easily solved by the use of
identity (\ref{conf top}). In order to see this, let us follow
\cite{rie} and introduce the conformal Green function
$G(x,x^\prime)$  of the operator $\De_D$, where
\beq
\sqrt{-g}\,\De_D^x\,G(x,x^\prime) \,=\,\de^D(x,x^\prime)
\,,\quad
G\,=\,{\bar G}\,.
\label{Green}
\eeq

The complete solution for the anomaly-induced effective
action can be written down in the form
\beq
\Ga_{ind}
&=&
S_c
\,+\,
\iint\limits_{x\,y} \Big\{
\frac{1}{4}\,c_r\,W^{r}_D
\,+\,
\frac{a}{8}\,{\tilde E}_D(x)\Big\}
\,G(x,y)\,{\tilde E}_D(y)
\nonumber
\\
&+&   % \limits
\sum_k \big( \ga_k - \al_k \big) \sum_i c_{ik}
\int_x{\cal L}_i\,.
\label{Ga}
\eeq
Here $\,S_c=S_c[g_{\mu\nu}]\,$ is an undefined conformal functional,
which represents a boundary condition of the variational equation
(\ref{indEA}), and the modification of the coefficients $\ga_k$ of
the anomaly (\ref{T}) occurs because part of the surface terms were
absorbed into ${\tilde E}_D$.

Writing the non-local part of the expression (\ref{Ga}) in the
symmetric form, one can always present the effective action in
the local covariant form which includes two auxiliary fields
$\psi$ and $\ph$, as it was suggested in \cite{a,MaMo}
\beq
{\bar \Ga} &=& S_c
\,+\,
\sum_k \big( \ga_k - \al_k \big) \sum_i c_{ik}
\int_x{\cal L}_i
\label{finaction}
\\
&+&
\frac12 \,
\int_x
\Big\{
\ph\De_D\ph - \psi\De_ D\psi
\,+\,\sqrt{-a}\,\ph\, {\tilde E}_D
\,+\, \frac{1}{\sqrt{-a}}\,(\psi-\ph)\,\,c_r\,W^{r}_D(x)\Big\}\,.
\nonumber
\eeq
In these formulas we assume that $a<0$, as in the $D=4$ case.
In case of $a> 0$ the expression can be trivially modified by
changing the sign $\,{\tilde E}_D \to -{\tilde E}_D$. The last
observation is that one can also write the action in terms of
modified auxiliary fields \cite{MaMo,P4} or in the simplest
non-covariant form in terms of $\si$ and ${\bar g}_{\mu\nu}$
\cite{rie}. Since the transition to these forms is not too different
compared to the $D=4$ case, we will not consider these issues here.

All in all, it is clear that the integration of anomaly needs
Eq.~(\ref{conf top}) at the first place and also Eq. (\ref{confLoc})
to deal with the local part of induced action. In the next section
we present the result for (\ref{conf top}) in $D=6$.

%%%%%%%%%%%%%%%%%%%%%%%%%%%
%%%%%%%%%%%%%%%%%%%%%%%%%%%
\section{Conformal formulas in $D=6$}
\label{main}

The candidate terms to the total derivatives in (\ref{T}) can be
reduced to the form \cite{FST}
\beq
\label{chi}
&&
\chi_1 = {\cx}^2{R}
\,, \quad
\chi_{2;3;4} \,=\,
{\cx}\big(
R^2_{\mu \nu \al \be};\, R^2_{\mu \nu};\, R^2\big)
\nonumber
\\
&&
\chi_{5;6;7;8}
\,=\, \na_{\mu}\na_{\nu}
\big(
R^{\mu}\,_{\la \al \be} R^{\nu \la \al \be}
;\,
R_{\al \be} R^{\mu \al \nu \be}
;\,
R_{\al}^{\mu} R^{\nu \al}
;\,
R R^{\mu \nu}\big)\,.
\eeq

After a very long and in fact complicated calculations, we arrived
at the following coefficients which guarantee the equations
(\ref{ED}) and (\ref{conf top}) for $D=6$,
\beq
\label{coef}
&&
\al_1 = \frac{3}{5}
\,,\quad
\al_2 = \frac{147}{20} +\xi
\,,\quad
\al_3 = -\frac{33}{5} -\frac{1}{2}\xi
\,,\quad
\al_4 = 0
\nonumber
\\
&&
\al_5 = -3+4\xi
\,,\quad
\al_6 = 6-4\xi
\,,\quad
\al_7 = 3-3\xi
\,,\quad
\al_8 = \xi\,.
\eeq
Here $\xi$ is a free parameter which remains undetermined
by the condition (\ref{conf top}). Assuming (\ref{coef}), all
the non-linear in $\si$ terms in (\ref{conf top}) cancel, and
the remaining linear term corresponds to $\ka = -6$ and the
conformal operator
\beq
\label{D6}
\De_6 &=&
{\cx}^3
+ 4R^{\mu \nu}\na_{\mu} \na_{\nu}{\cx}
- R{\cx}^2
\nonumber
\\
&+&
4\na_{\al}\big[(\na^{\al}R^{\mu \nu}) \na_{\mu} \na_{\nu}\big]
+ V^{\mu\nu}\na_{\mu} \na_{\nu} + N^\la \na_\la\,,
\eeq
where
\beq
V^{\mu\nu}
&=&
\frac{12+4\xi}{3}\big( R_{\al\be} R^{\mu \al \nu \be}
-  R^{\mu}\,_{\al \be \ga}R^{\nu \al \be \ga}\big)
\,+\,
(9+\xi) \Big(
R^{\mu \al} R^{\nu}_{\al}
- \frac{1}{3}R R^{\mu \nu}\Big)
\nonumber
\\
&+&
g^{\mu\nu}\Big[
\frac{81 + 20 \xi}{15} R^2_{\mu \nu \al \be}
- \frac{69 + 15\xi}{10} R^2_{\mu \nu}
+ \frac{6+\xi}{6} R^2
- \frac{3}{5}({\cx}R)    \Big]
\nonumber
\eeq
and
\beq
N^\la
&=&
\frac{44+10\xi}{5}
R_{\mu \nu \al \be}( \na^\la R^{\mu \nu \al \be})
+ \frac{12+4\xi}{3} R^{\mu \nu \al \la}( \na_{\mu}
R_{\nu \al})
- \Big(\frac{49}{5}+\frac{5\xi}{3}\Big)
R_{\mu \nu}(\na^\la R^{\mu \nu})
\nonumber
\\
&+&
\frac{15-\xi}{3}R_{\mu \nu}(\na^{\mu}R^{\nu \la})
+ \frac{9+\xi}{6} R^{\mu \la}(\na_{\mu}R)
+  \frac{3+\xi}{6} R (\na^\la R)
+ \frac{2}{5}(\na^\la{\cx}R)\,.
\nonumber
\eeq
One has to remember that here the covariant derivative does
not act beyond the parenthesis.

Let us note that in the literature one can find a general theory for
constructing conformal operators (see, e.g.,
\cite{Conf3,Arak,Hamada,Osborn}),
still the operator (\ref{D6}) is more general that the ones known
before. The main relation (\ref{ED}) was not derived before,
probably  due to the complexity of calculations requested to get
the coefficients (\ref{coef}). We could achieve it by combining
hand-made work and the softwares Cadabra \cite{Cadabra} and
Mathematica \cite{Wolfram}.  The essential details will be
published elsewhere \cite{FF-big}, together with the solution
for the local terms producing surface terms (\ref{confLoc}) in
the anomaly.

Compared to the main calculation, it is much easier (but still
consuming certain time and effort) to check that the operator
$\De_6$ satisfies the conformal invariance (\ref{confOP}) and is
self-adjoint, $\,\int_x \ph \De_6 \chi =  \int_x \chi \De_6 \ph$.
It is interesting that both conditions do not pose any restriction
on the value of an arbitrary parameter $\xi$. We shall discuss
the physical consequence of this ambiguity in the last section.

%%%%%%%%%%%%%%%%%%%%%%%%%%%
\section{Conclusions and discussions}
\label{Conc}

The equations  (\ref{coef}) and  (\ref{D6}) form the full set of the
building blocks for the non-local part of anomaly-induced action
(\ref{finaction}) in $D=6$. Together with the previously known
examples in $D=2,4$ this enables us to draw some general
conclusions and discuss the similarities and differences between
the new result and the previous one. One of the common points is
that the anomaly-induced expression is an exact effective action
for the homogeneous and isotropic metric, where the conformal
functional $S_c$ is irrelevant. Assuming that the space-time has
six dimensions, and that there are massless conformal fields in
the far IR, we arrive at the exact solution for anomaly-induced
action in this particular class of metrics.

Qualitatively, the structure of (\ref{Ga}) and  (\ref{finaction})
is the same in all even dimensions, but the complexity of the
solution increases with dimension. On the transition from two
to four dimensions the main complications were the integration
constant $S_c$ and the presence of the two different (conformal
and topological) terms in (\ref{T}) which produce non-local terms
in the anomaly-induced action \cite{DDI}. One of the consequences
is that the integrated anomaly can be  consistently written in local
covariant form only by means of two auxiliary fields
\cite{a,MaMo,PoImpo},
while in $D=2$ one such field is sufficient. As we have seen in
Sect.~\ref{gen} the number of auxiliary fields remains the same
in higher dimensions. At the same time the solution (\ref{coef}),
(\ref{D6}) includes a qualitatively new arbitrary parameter $\xi$.
Nothing of this sort takes place in  $D=2,4$. An interesting
possibility is that the ambiguity can be fixed by imposing the
consistency conditions \cite{Bonora,Bastianelli,Grinshtein}, but
it is not certain, of course. Another question is what could be
the physical effect of an arbitrary parameter $\xi$?

Since the conformal anomaly is the same for any $\xi$, one can
simply ignore the ambiguity by fixing some particular value for
this parameter. The difference between distinct values can be always
absorbed into the conformal functional $S_c$. The situation is
technically similar to the one with the $\psi$-dependent part of
(\ref{finaction}), which can be also absorbed into conformal part.
However, in the case of $\psi$-terms this would be a wrong idea.
For instance, without the second auxiliary field one can not classify
vacuum states in the vicinity of the spherically symmetric black
holes \cite{balsan}. There is no such a problem for the gravitational
waves, but maybe only because all known calculations were done
for the isotropic cosmological backgrounds
\cite{star83,wave,HHR,GW-Stab}. Concerning the role of $\xi$,
the question is whether it affects the relevant solutions, and this
question will remain open until such solutions are explored for
the action  (\ref{finaction}).

The last observation concerns the possible applications of the
effective actions  (\ref{Ga}) and (\ref{finaction}). One can imagine
that the explicit form of effective vacuum action for the conformal
fields can be useful for verifying the calculations related to
holography and AdS/CFT correspondence.  Another application
is related to the dimensional reduction to $D=4$, expected to
produce a four-dimensional action different from the one coming
from integrating anomaly directly in  $D=4$. Due to the universality
of the result, the calculation of such a reduced action and the study
of its physically relevant solutions may be eventually useful in
designing the experimental and/or observational tests for the
existence of extra dimensions.

%%%%%%%%%%%%%%%%%%%%%%%%%%%%%%%
%%%%%%%%%%%%%%%%%%%%%%%%%%%%%%%
\section*{Acknowledgements}
I.Sh. is grateful to CNPq, FAPEMIG and ICTP for partial
support.
%%   and to the Mainz Institute for Theoretical Physics
%%  (MITP) for its hospitality and partial
%%   support during the completion of the first version
%%   of this work.

%%%%%%%%%%%%%%%%%%%%%%%%%%%%%%%
%%%%%%%%%%%%%%%%%%%%%%%%%%%%%%%
%%%%%%%%%%%%%%%%%%%%%%%%%%%%%%%

\end{document}